\documentclass[pra,aps,twocolumn,nofootinbib]{revtex4}%
\usepackage{hyperref}
\usepackage{amsmath}
\usepackage{amsfonts}
\usepackage{amssymb}
\usepackage{graphicx}%
\begin{document}
\title{Efficient compression of quantum information}
\author{Martin Plesch$^{1,2}$ and Vladimir Buzek$^{1}$ }
\affiliation{${}^{1}$ Institute of Physics, Slovak Academy of
Sciences, D\'{u}bravsk\'{a} cesta 9, 845 11 Bratislava, Slovakia }
\affiliation{${}^{2}$ Faculty of Physics, University of Vienna,
Boltzmanngasse 5, Vienna, Austria}
\date{14 January 2010}

\begin{abstract}
We propose a scheme for an exact efficient transformation of a
tensor product state of many identical qubits into a state of an
exponentially small number of qubits. Using a quadratic number of
elementary quantum gates we transform $N$ identically prepared
qubits into a state, which is nontrivial only on the first
$\left\lceil \log(N+1)\right\rceil$ qubits. This procedure might
be useful for quantum memories, as only a small portion of the
original qubits has to be stored. A second possible application is
in communicating a direction encoded in a set of quantum states,
as the compressed state provides a high-effective method for such
an encoding.
\end{abstract}
\maketitle

\section{Introduction} Product states of many identical copies of
a one-qubit state are a specific type of symmetric states. Having
only two parameters, they span the symmetric subspace with linear
dimension $N+1$ (when $N$ is the number of the copies). On the
other hand, this subspace is exponentially small in comparison to
the whole Hilbert space of all qubits, which has dimension $2^N$.
Thus, one may ask if (and how) it would be possible to
``compress'' information encoded in an $N$-fold product state of a
single qubit state into a smaller number of qubits, prepared in a
complicated, possibly entangled state. Comparing the dimensions of
the Hilbert space of symmetric states of $N$ qubits ($N+1$) with
the whole Hilbert space of a smaller number $n$ of qubits ($2^n$)
one can immediately see that the number of qubits needed to store
the compressed state is $n=\left\lceil \log(N+1)\right\rceil$.

Gisin and Popescu \cite{Gisin} showed that two qubits in
antiparallel states provide a better encoding of a direction than
two copies of the same qubit. In a sense, one might see even these
two antiparallel spins as a compressed state, representing a
higher (though not natural) number of copies of a single qubit. In
\cite{Massar} it was proved that sending of a direction of one
qubit is optimally performed by sending two antiparallel states.
The proof is relying on the fact that the sender and receiver
should not share a common reference frame. More general research
on this topic was performed later in \cite{Bagan}.

However, if we relax the condition of not sharing a reference
frame between communicating parties, it is expectable that we can
communicate the direction in a more effective way. In this case
the possible encoding and decoding procedures may include
basis-dependent operations and thus allow for a more effective
communication. A possible scenario is to compress identical
one-qubit states (pointing into the desired direction) and
communicate only the compressed state. The other party can
decompress the state and perform state-tomography on an
exponentially higher number of qubits.

An other possible scenario for utilizing the compression procedure
is a quantum memory. Both the encoding and the decoding will be
done by the same party, so the correct reference frame will always
be available. Having a-priori information about the fact that a
set of qubits is prepared in a symmetric state, we can reduce the
resources needed by storing just the compressed state.

However, any compression algorithm\footnote{The suggested scheme
should not be confused with the Schumacher compression
\cite{Nielsen}. This compression is suitable for known quantum
sources, whereas our scheme is designed for unknown sources.} will
be of possible practical use only in the case it can be performed
in reasonable time, using reasonable resources. Such a condition
is usually understood as performing at most a polynomial number of
elementary (local) operations with respect to the number of
qubits. If we allow a small error $\epsilon$ in the compressing
operation, then methods to design circuits to perform the Schur
transform are known even for qudits \cite{Schur}. These circuits
are polynomial in the dimension $d$ of the qudits, the number $N$
of qudits and $\log(\epsilon^{-1})$.

The situation changes if we insist on performing the unitary
transformation exactly, not allowing any errors. In this case we
cannot utilize the Solovay-Kitaev theorem \cite{Nielsen}, which
implies the existence of effective quantum circuits, containing
operations only from a discrete set, and approximating any unitary
in an effective way. Instead of this, we will work with the
standard gate library \cite{Kniznica}, consisting of the Control
NOT gate (as a single two-qubit gate) and a continuous set of
single-qubit gates. With gates from this library, it is possible
to exactly perform any unitary transformation. However this
requires in general an exponential number of gates to be used.
Contrary to the general case, our circuit uses only a polynomial
(quadratic) number of elementary gates.

In the scenario of using the compression procedure for quantum
memories the fact of not having classical information about the
state of the qubits is important. If one only knows the fact that
a set of qubits is prepared in a separable symmetric state (i.e.
all qubits are in identical state) without any classical knowledge
about the state of individual qubits itself, unitary operations
have to be used to compress (and decompress) the overall state.
Contrary, knowing the state of the qubits classically, one is able
to calculate the amplitudes of the compressed state classically
and prepare the state directly on $n=\left\lceil
\log(N+1)\right\rceil$ qubits.

For decompression procedure, which is just the inverse operation
of the compression, the assumption of not having the classical
information about the compressed state is well justified in both
scenarios. In case of sending a direction the set of qubits sent
shall be the sole resource available (except of shared reference
frame); the same holds for quantum memory.

Similar research was performed by Phillip Kaye and Michele Mosca.
In \cite{Kaye1} they suggest an algorithm for effective
entanglement concentration. However, before applying their
algorithm, they perform a POVM on their states. Such method is
competent in cases, where we wish to utilize only some quality of
the states (say entanglement), but is not suitable if we need to
store all of the parameters of the unknown state. In \cite{Kaye2},
the authors suggest an effective algorithm for preparation of
(classically) known states, which is a conceptually different
problem, leading to a different solution.

The paper is organized as follows: in Section II we define
symmetric states and computational states, which are specific
states written in the computational basis. In Section III we
describe the transformation procedure of symmetric states into
computational states, including an example for three qubits. In
Section IV we describe the final procedure, which transforms
computational states into states non-trivially occupying only the
subspace of the first $\left\lceil \log(N+1)\right\rceil$ qubits.
Finally, in Section V we discuss possible further optimization of
the scheme and suggest possible applications.


\section{Symmetric states }Any symmetric state of $N$ qubits
exhibits the property
\begin{align}
\left\vert \Psi\right\rangle _{123...N}  &  =\left\vert
\Psi\right\rangle _{\sigma(123...N)}, ] \label{stavPSI}
\end{align}
where $\sigma(.)$ denotes a permutation of the individual qubit
systems. A basis for the set of symmetric states can be chosen so,
that every basis state has a definite number of excitations
(qubits in the state $\left\vert 1\right\rangle $) and respective
basis states can be
labelled by this number%
\begin{eqnarray}
\left\vert N;k\right\rangle =\binom{N}{k}^{-\frac{1}{2}}%
{\displaystyle\sum\limits_{\sigma}}
\sigma\left(  \left\vert 1\right\rangle ^{\otimes
k}\otimes\left\vert
0\right\rangle ^{\otimes(N-k)}\right)  . \label{Baza}%
\end{eqnarray}
The basis states are perpendicular to each other and normalized
\begin{equation}
\left\vert \langle N;k\left\vert N;l\right\rangle \right\vert
=\delta_{kl},
\label{Norma}%
\end{equation}
where the sum runs through all permutations of the qubit systems,
having $\binom{N}{k}$ terms. We suggest a transformation which
takes the symmetric states (\ref{Baza}) into a subset of
computational basis vectors. This subset is formed by the vector
$\left\vert 0\right\rangle ^{\otimes N}$ and all vectors having a
single excitation. It occupies the Hilbert space of the same
dimension as symmetric states and is defined as
\begin{align} \left\vert C\right\rangle
_{k}  &  =\left\vert 0\right\rangle ^{\otimes
(k-1)}\otimes\left\vert 1\right\rangle \otimes\left\vert
0\right\rangle
^{\otimes(N-k)}\label{Nova_baza}\\
\left\vert C\right\rangle _{0}  &  =\left\vert 0\right\rangle
^{\otimes N}.\nonumber
\end{align}

This subset is very accessible for the computation for two
reasons:
\begin{itemize}
\item It is easy to change a state, as only a two qubit operation
is needed to take one basis state to an other one.

\item It acts as a control very easy, as every basis state is
defined just by a position of a single excitation, which can act
as a control qubit.
\end{itemize}


\section{Transformation }We suggest a transformation $U$ in the form%
\begin{equation}
U\left(\left\vert N;k\right\rangle\right) =\left\vert
C\right\rangle _{k}.
\label{Transformacia}%
\end{equation}
This transformation is not defined on the whole Hilbert space,
which leaves some possibilities for further optimization. However,
even without any optimization we will show that it is possible to
implement (\ref{Transformacia}) with $O(N^{2})$ elementary gates.
Let us examine the cases of few qubits first.

\subsection{One qubit }For one qubit the situation is rather
trivial and no transformation is needed,
\begin{align} \left\vert
0\right\rangle  &  \longrightarrow\left\vert 0\right\rangle
\label{One_qubit}\\
\left\vert 1\right\rangle  &  \longrightarrow\left\vert
1\right\rangle .\nonumber
\end{align}

\subsection{Two qubits }Here we need to perform a transformation
only on a part of the whole Hilbert space:%
\begin{align}
\left\vert 00\right\rangle & \longrightarrow\left\vert
00\right\rangle
\label{Two_qubits}\\
\left\vert 01\right\rangle +\left\vert 10\right\rangle
& \longrightarrow \sqrt{2}\left\vert 10\right\rangle \nonumber\\
\left\vert 11\right\rangle & \longrightarrow\left\vert
01\right\rangle. \nonumber
\end{align}

In the second row of (\ref{Two_qubits}) the symmetric combination
of two states possessing a single excitation is combined to the
state $\left\vert 10\right\rangle $. The state $\left\vert
1\right\rangle $ is on the first position, encoding a single
excitation of the original state. In the third row the state
$\left\vert 11\right\rangle $ is transformed into $\left\vert
01\right\rangle $, encoding two original excitations into
excitation on the second position.

For two qubits, only a single state is not defined by this
transformation allowing one parameter for further optimization
\begin{equation}
\frac{1}{\sqrt{2}}\left( \left\vert 01\right\rangle -\left\vert
01\right\rangle \right) \longrightarrow e^{i\phi}\left\vert
11\right\rangle .
\end{equation}
 In general (as a two qubit operation) it is
realizable by at most three CNOT gates in combination with
single-qubit operations.

\subsection{Three qubits}From eight independent basis states of
the three qubits
Hilbert space the  operation $U$ defines only four states:%
\begin{align}
\left\vert 000\right\rangle & \longrightarrow\left\vert
000\right\rangle
\label{Three_qubits}\\
\frac{1}{\sqrt{3}}\left(  \left\vert 001\right\rangle +\left\vert
010\right\rangle +\left\vert 100\right\rangle \right)   &
\longrightarrow
\left\vert 100\right\rangle \nonumber\\
\frac{1}{\sqrt{3}}\left(  \left\vert 011\right\rangle +\left\vert
101\right\rangle +\left\vert 110\right\rangle \right)   &
\longrightarrow \left\vert 010\right\rangle \nonumber\\
\left\vert 111\right\rangle  & \longrightarrow\left\vert
001\right\rangle \nonumber
\end{align}
Similar to the case of two qubits, there is a simple logic behind
this operation. We need to combine all states having the same
number of excitations, taken with equal weights and equal phases,
into one single state with a single excitation on the proper
position. This can be clearly seen in the second and third row of
the definition (\ref{Three_qubits}).

In this case there are four more basis states, for which the
operation is undefined, leaving us with $12$ free parameters. Even
without utilization of this option one needs at most $22$ C-NOT
gates to perform (any) three-qubit operation \cite{Cosin}.

\subsection{More qubits. }For more qubits, the number of C-NOT
gates needed to perform a general operation grows exponentially
and is not known exactly. Attempts to perform a general
optimizations have been made in several papers
\cite{Cosin,Optimalizacia,Sedlak} with only partial success. Here
we suggest a sequence of small (three qubit) operations, which
follows the logic displayed on the two and three qubit cases and
guarantees a quadratic number of C-NOT gates and local operations
with respect to the number of qubits. Moreover, the free
parameters in operations used allow further optimization of this
scheme.

We will define the scheme on the basis states of symmetric
subspace of the $N $-qubit Hilbert space. Due to linearity, if the
scheme performs operation $U$ on basis states, it does so on any
symmetric state. For non-symmetric states (which occupy the
substantial portion of Hilbert space of many qubits) the action of
the operation may be arbitrary.

Let us start with a basis state $\left\vert N;k\right\rangle $.
The number of qubits $N$ is supposed to be known and the operation
may and will depend on it. On the contrary, the number of
excitations $k$ must not be part of the definition of the
operation itself, as the operation is applied on a superposition
of states with a fixed $N$, but different $k$s.

As the first step we perform the operation (\ref{Two_qubits}) on
the first two
qubits of the state:%
\begin{align}
\left\vert N;k\right\rangle  &  =\binom{N}{k}^{-\frac{1}{2}}%
{\displaystyle\sum\limits_{\sigma}}
\sigma\left(  \left\vert 1\right\rangle ^{\otimes
k}\otimes\left\vert
0\right\rangle ^{\otimes(N-k)}\right) \label{Op_1}\\
&  \longrightarrow\left\vert 00\right\rangle \binom{N-2}{k}^{-\frac{1}{2}}%
{\displaystyle\sum\limits_{\binom{N-2}{k}}}
P\left(  \left\vert 1\right\rangle ^{\otimes k}\otimes\left\vert
0\right\rangle ^{\otimes(N-2-k)}\right) \nonumber\\
&  +\sqrt{2}\left\vert 10\right\rangle \binom{N-2}{k-1}^{-\frac{1}{2}}%
{\displaystyle\sum\limits_{\binom{N-2}{k-1}}}
P\left(  \left\vert 1\right\rangle ^{\otimes k-1}\otimes\left\vert
0\right\rangle ^{\otimes(N-1-k)}\right) \nonumber\\
&  +\left\vert 01\right\rangle \binom{N-2}{k-2}^{-\frac{1}{2}}%
{\displaystyle\sum\limits_{\binom{N-2}{k-2}}}
P\left(  \left\vert 1\right\rangle ^{\otimes k-2}\otimes\left\vert
0\right\rangle ^{\otimes(N-k)}\right)  .\nonumber
\end{align}
For this operation one needs no more than three C-NOT gates. The
$\sqrt{2}$ in the third row of the definition (\ref{Op_1}) comes
from the fact that the state beginning with $\left\vert
10\right\rangle $ contains two original states (both beginning
with $\left\vert 10\right\rangle $ and $\left\vert 01\right\rangle
$).

Now we have virtually divided the state of $N$ qubits into two
parts. In the first part (two qubits) the logic of the output
basis is implemented, where the position of the excitation encodes
the number of excitations originally contained in the first part
of the state. The second part of the state is in its original
form, symmetric with respect to the permutation of qubits within
this part.

We will proceed with the transformation to gradually enlarge the
transformed part of the state. To do this, we will take the first
qubit (let us denote this qubit as the $a$th qubit) of the
non-transformed part of the state. We will perform specific three
qubit operations on this qubit and any neighboring pair of qubits
in the transformed part of the state. This operations will perform
following actions:

\begin{enumerate}
\item If the $a$th qubit is in the state $\left\vert
0\right\rangle $, no change needs to be done to the transformed
part of the state, as the excitation is on the proper position
also including the $a$th qubit into the transformed part of the
state

\item If the $a$th qubit is in the state $\left\vert
1\right\rangle $, the sequence of operations will "scan" the
transformed state and shift the excitation by one position to the
right and remove the excitation from the $a$th qubit

\item Specifically, if the $a$th qubit is in the state $\left\vert
1\right\rangle $ and there was no excitation so far in the
transformed part of the state, the operation will switch the first
qubit to the state $\left\vert 1\right\rangle $ and remove the
excitation from the $a$th qubit at the same time

\item Specifically, if the $a$th qubit is in the state $\left\vert
1\right\rangle $ and the excitation in the transformed part of the
string is on the last position (qubit $a-1$), the operation will
remove this excitation, but will keep the excitation on the $a$th
qubit.
\end{enumerate}

Written in mathematical terms, omitting the part of the state
starting with
the qubit $a+1$, we will perform the operation $U(a)$ as follows:%
\begin{align}
\left\vert \psi\right\rangle \left\vert 0\right\rangle _{a}  &
\longrightarrow\left\vert \psi\right\rangle \left\vert
0\right\rangle
_{a}\label{Op_2}\\
\left\vert 0...0\right\rangle \left\vert 1\right\rangle
_{b}\left\vert 0...0\right\rangle \left\vert 1\right\rangle _{a} &
\longrightarrow \left\vert 0...0\right\rangle \left\vert
1\right\rangle _{b+1}\left\vert
0...0\right\rangle \left\vert 0\right\rangle _{a}\nonumber\\
\left\vert 0...0\right\rangle \left\vert 1\right\rangle _{a}  &
\longrightarrow\left\vert 1\right\rangle \left\vert
0...0\right\rangle
\left\vert 0\right\rangle _{a}\nonumber\\
\left\vert 0...0\right\rangle \left\vert 1\right\rangle \left\vert
1\right\rangle _{a}  &  \longrightarrow\left\vert
0...0\right\rangle \left\vert 1\right\rangle _{a}.\nonumber
\end{align}

To perform this transformation, we need to apply a three qubit
operation $U(a,b)$ on qubits on the positions $b$, $b+1$ and $a$,
for every $b$ running from $1$ to
$a-2$:%
\begin{align}
\left\vert 00\right\rangle _{b}\left\vert 0\right\rangle _{a}  &
\longrightarrow\left\vert 00\right\rangle _{b}\left\vert
0\right\rangle
_{a}\label{Op_3}\\
\left\vert 10\right\rangle _{b}\left\vert 0\right\rangle _{a}  &
\longrightarrow\left\vert 10\right\rangle _{b}\left\vert
0\right\rangle
_{a}\nonumber\\
\left\vert 00\right\rangle _{b}\left\vert 1\right\rangle _{a}  &
\longrightarrow\left\vert 00\right\rangle _{b}\left\vert
1\right\rangle
_{a}\nonumber\\
\left\vert 01\right\rangle _{b}\left\vert 1\right\rangle _{a}  &
\longrightarrow\left\vert 01\right\rangle _{b}\left\vert
1\right\rangle
_{a}\nonumber\\
\alpha_{101}\left\vert 10\right\rangle _{b}\left\vert
1\right\rangle _{a}+\alpha_{010}\left\vert 01\right\rangle
_{b}\left\vert 0\right\rangle _{a}  &
\longrightarrow\beta_{010}\left\vert 01\right\rangle
_{b}\left\vert 0\right\rangle _{a},\nonumber
\end{align}
where
\begin{align}
\alpha_{101}=\sqrt{\binom{a-1}{b}}\nonumber\\
 \alpha_{010}=\sqrt{\binom{a-1}{b+1}
}\nonumber\\
\beta_{010}=\sqrt{\binom{a}{b+1}}\nonumber
\end{align}
and
\begin{equation}
\left\vert 00\right\rangle _{b}=\left\vert 0\right\rangle
_{b}\left\vert 0\right\rangle_{b+1}.
\end{equation}

 The first two rows of the operation (\ref{Op_3})
obey the first condition posed on the transformation - if the
$a$th qubit is not excited, the string should not be changed. The
third and fourth row are part of the "scanning" process, where we
need to find the excitation in the transformed string and push it
by one position. In the third row we did not find the excitation,
so no action is performed. In the fourth row the excitation was
found, but should be transformed to the position $b+2$, which is
not part of the transformation, so here is no action required
again. The crucial part of the transformation is in the fifth row.

The state $\left\vert 01\right\rangle _{b}\left\vert
0\right\rangle _{a}$ should not be transformed obeying the first
condition, as the state of the $a$th qubit is $\left\vert
0\right\rangle$. However, the state $\left\vert 10\right\rangle
_{b}\left\vert 1\right\rangle _{a}$ should be transformed to
$\left\vert 01\right\rangle _{b}\left\vert 0\right\rangle _{a} $
obeying the second condition. This can not be done separately, as
this would induce a non unitary operation (two perpendicular
states would result into two identical states). What can be done
is to transform a specific linear combination of these two states.

Let us change the normalization till the end of this section and
suppose that all states that formed the original state $\left\vert
N;k\right\rangle $ (written in computational basis) had norm $1$
(this would result in the norm $\binom{N}{k}$ of the state
$\left\vert N;k\right\rangle$). Then the partially transformed
state containing $\left\vert 10\right\rangle
_{b}\left\vert 1\right\rangle _{a}$ will have the amplitude $\sqrt{\binom{a-1}{b}}%
$, which comes from the fact that there are already combined all
states which contained $b$ excitations within $a-1$ positions. The
same holds for the state $\left\vert 01\right\rangle
_{b}\left\vert 0\right\rangle _{a}$, where the amplitude is
$\sqrt{\binom{a-1}{b+1}}$. For the state $\left\vert
01\right\rangle _{b}\left\vert 0\right\rangle _{a}$ after
transformation the amplitude is $\sqrt{\binom{a}{b+1}}$, as we
have $b+1$ excitations within $a$ qubits. Preservation of the norm
by the transformation can be seen very easily, taking the squares
of amplitudes we get combinatorial numbers forming a small
edge-down triangle in the Pascal triangle, where a rule applies
that the number on a
specific position is given by the sum of two numbers above it, e.g.%
\begin{equation}
\binom{a}{b+1}=\binom{a-1}{b}+\binom{a-1}{b+1}. \label{Op_5}%
\end{equation}

To successfully conclude the operation $U(a)$ (\ref{Op_2}) for a
specific $a$, we still need to apply the last two conditions,
dealing with the specific cases of $0$ and $a$ excitations in the
transformed string. To do that, we will perform an operation
acting on the first qubit and on the pair of qubits
on the positions $a-1$ and $a\,$:%
\begin{align}
\left\vert 0\right\rangle _{1}\left\vert 00\right\rangle _{a-1}  &
\longrightarrow\left\vert 0\right\rangle _{1}\left\vert
00\right\rangle
_{a-1}\label{Op_6}\\
\left\vert 0\right\rangle _{1}\left\vert 10\right\rangle _{a-1}  &
\longrightarrow\left\vert 0\right\rangle _{1}\left\vert
10\right\rangle
_{a-1}\nonumber\\
\left\vert 0\right\rangle _{1}\left\vert 11\right\rangle _{a-1}  &
\longrightarrow\left\vert 0\right\rangle _{1}\left\vert
01\right\rangle
_{a-1}\nonumber\\
\alpha_{001}\left\vert 0\right\rangle _{1}\left\vert
01\right\rangle _{a-1} &  +\alpha_{100}\left\vert 1\right\rangle
_{1}\left\vert 00\right\rangle
_{a-1}\longrightarrow\beta_{100}\left\vert 1\right\rangle
_{1}\left\vert 00\right\rangle _{a-1},\nonumber
\end{align}
where%
\begin{equation}
\alpha_{001}=1;  \alpha_{100}=\sqrt{a-1};
\beta_{100}=\sqrt{a}.\nonumber
\end{equation}
Here the first two rows of the operation obey the first condition
that for no excitation on the $a$th position no action is
required. The third row applies the fourth condition; if $a-1$
excitations were in the original non-transformed state (resulting
in the excitation of the position $a-1$ in the transformed state)
and $a$th qubit is excited, it should remain excited but the
excitation of the qubit on the position $a-1$ has to be removed.
The last row of (\ref{Op_6}) similarly to the situation in
(\ref{Op_3}) combines two states in a specific superposition. The
state $\left\vert 0\right\rangle _{1}\left\vert 01\right\rangle
_{a}$ has a unit norm, as it was not combined till now with any
other state. State $\left\vert 1\right\rangle _{1}\left\vert
00\right\rangle _{a}$ before transformation has the amplitude
$\sqrt{a-1}$ (one excitation among $a-1$ possible positions) and
the state $\left\vert 1\right\rangle _{1}\left\vert
00\right\rangle _{a}$ after transformation has the amplitude
$\sqrt{a}$ (one excitation among $a$ possible positions).

For every $a$ from $3$ to $N$ we have to perform $a-2$ operations
of the type (\ref{Op_2}) and one operation of the type
(\ref{Op_6}). This results in altogether
\begin{equation}
{\displaystyle\sum\limits_{3}^{N}} \left(  a-2\right)  +\left(
N-2\right)  =\frac{(N+1)(N-2)}{2}
\end{equation}
 three-qubit operations, plus a
two-qubit operation from the very first step. As any three-qubit
operation can be realized by at most $21$ C-NOT gates (plus local
transformations) and any two-qubit operation by at most $3$ CNOT
gates (plus local transformations), we get as the upper bound
\begin{equation}
n(N)=\frac{21}{2}\left( N^{2}-N-2\right)  +3,
\end{equation}
a quadratic dependence on the number of qubits. This is far better
than any optimization method can perform in a general case and
causes an exponential speed-up in comparison to any known general
decomposition. Moreover, the open parameters in the definition of
the operations (\ref{Op_3}) and (\ref{Op_6}) may allow for further
optimization. Also optimization of the final configuration may
result in further decrease of the number of CNOTs needed, however
most probably by keeping the quadratic dependence on the number of
qubits.

\subsection{Five qubits example} As the above described procedure
is rather complicated and not easy to understand, we present an
example of five qubits. In this case, the input state has the form%
\begin{align}
\left\vert \Psi\right\rangle  & =\left\vert \psi\right\rangle
^{\otimes 5}=\left(  \alpha\left\vert 0\right\rangle
+\beta\left\vert 1\right\rangle
\right)  ^{\otimes5}\label{five_one}\\
& =\alpha^{5}\left\vert 00000\right\rangle
+\sqrt{5}\alpha^{4}\beta\left\vert
5;1\right\rangle \nonumber\\
& +\sqrt{10}\alpha^{3}\beta^{2}\left\vert 5;2\right\rangle +\sqrt{10}%
\alpha^{2}\beta^{3}\left\vert 5;3\right\rangle \nonumber\\
& +\sqrt{5}\alpha\beta^{4}\left\vert 5;4\right\rangle
+\beta^{5}\left\vert
11111\right\rangle. \nonumber%
\end{align}

Let us now apply the transformation step by step on one of
components of the state (\ref{five_one}), e.g. on $\left\vert
5;3\right\rangle $. In further steps we omit the amplitude of the
state in the original state $\left\vert \Psi\right\rangle $ given
by $\alpha$ and $\beta$, but keep the norm factor $\sqrt{10}$ for
simplicity. As the first operation we apply (\ref{Op_1}) on the
first
two qubits. This results in the state%
\begin{equation}
\sqrt{10}\left\vert 5;3\right\rangle \rightarrow\left\vert
00\right\rangle \left\vert 111\right\rangle +\sqrt{6}\left\vert
10\right\rangle \left\vert 3;2\right\rangle +\sqrt{3}\left\vert
01\right\rangle \left\vert
3;1\right\rangle. \label{five_two}%
\end{equation}

\begin{figure}
\includegraphics[width=\hsize]{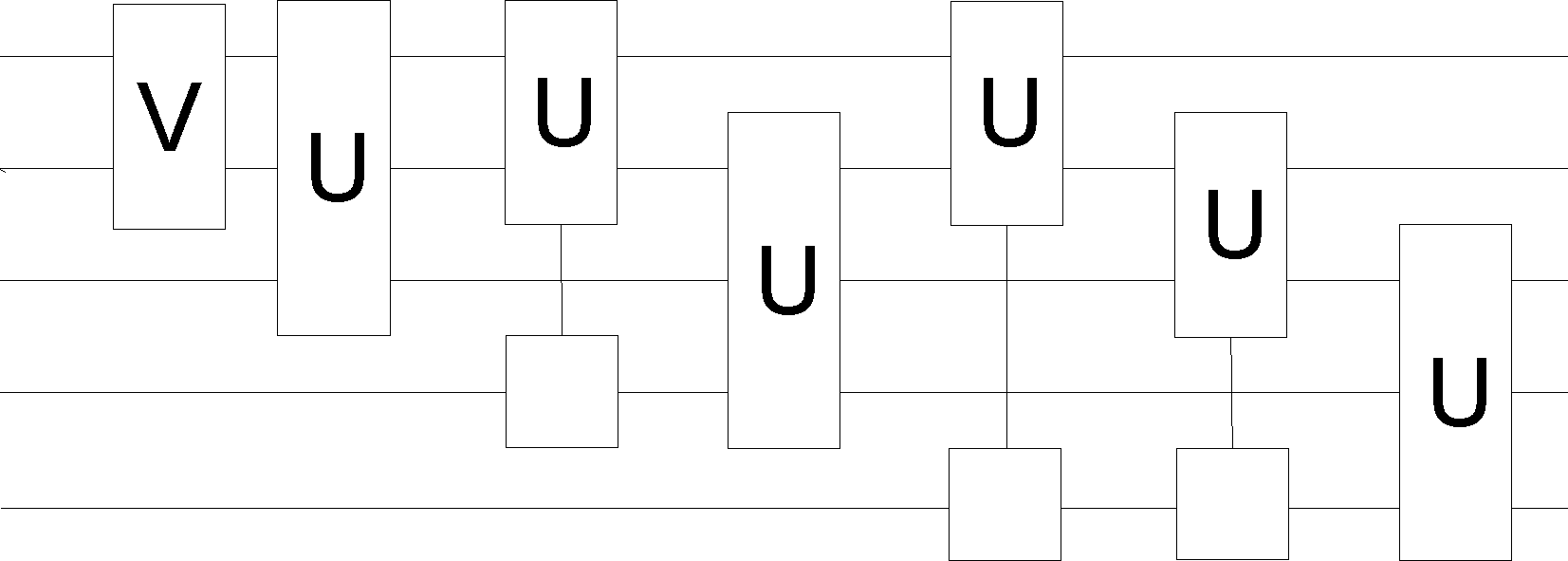}
\caption{ \label{figure1} Sequence of gates of compression
transformation in the five-qubits example. Operation $V$
represents the starting two-qubit transformation (\ref{Op_1}) and
operations $U$ represent relevant $U(a,b)$ operations
(\ref{Op_3}).}
\end{figure}

Now we apply the operation (\ref{Op_3}); Indices $a$ and $b$ run
from $3$ to $5$ and from $1$ to $a-2$ respectively. Graphical
representation of the circuit is depicted in the Figure
\ref{figure1} and results of the operations after each step are
shown in the Table \ref{table_five}.
\begin{table}
\begin{center}
\begin{tabular}
[c]{|l|l|l|} \hline $a$ & $b$ & Result after
transformation\\
\hline
$3$ & $1$ & $\sqrt{3}\left\vert
100\right\rangle \left\vert 11\right\rangle +\sqrt{6}\left\vert
010\right\rangle \left\vert 2;1\right\rangle +\left\vert
001\right\rangle \left\vert 00\right\rangle $\\
\hline
 $4$ & $1$ & $\sqrt{6}\left\vert 0100\right\rangle
\left\vert 1\right\rangle +\sqrt{3}\left\vert 010\right\rangle
\left\vert 10\right\rangle +\left\vert
001\right\rangle \left\vert 00\right\rangle $\\
\hline
 $4$ & $2$ & $\sqrt{6}\left\vert 0100\right\rangle
\left\vert 1\right\rangle
+\sqrt{4}\left\vert 0010\right\rangle \left\vert 0\right\rangle $\\
\hline
 $5$ & $1$ & $\sqrt{6}\left\vert 0100\right\rangle
\left\vert 1\right\rangle
+\sqrt{4}\left\vert 0010\right\rangle \left\vert 0\right\rangle $\\
\hline
$5$ & $2$ & $\sqrt{10}\left\vert 00100\right\rangle $\\
\hline $5$ & $3$ & $\sqrt{10}\left\vert 00100\right\rangle $\\
\hline
\end{tabular}
\end{center}
\caption{The resulting state after partial transformation $U(a,b)$
is displayed for the special case of transformation of the state
$\left\vert 5;3\right\rangle$.} \label{table_five}
\end{table}

The state $\sqrt{10}\left\vert 5;3\right\rangle $ was transformed
to the state $\sqrt{10}\left\vert 00100\right\rangle $, i.e. the
number of excitations in the state was transformed into the
position of a single excitation. In every step of the operation
(in the state (\ref{five_two}) and in every row of the Table
\ref{table_five}) the \textbf{position} of the excitation in the
"processed" part of the state (denoted as the first ket) plus the
\textbf{number} of excitations in the "unprocessed" part of the
system (denoted as the second ket) sum to three, the number of
excitations in the untransformed state.

\section{Final step }As a final step of the procedure, we need to
perform a transformation
\begin{equation}
\left\vert C\right\rangle _{k}\longrightarrow\left\vert
B\right\rangle _{k},
\end{equation}
where $\left\vert B\right\rangle _{k}$ is a set of $N+1$ states
occupying nontrivially only the subspace of $\left\lceil
\log(N+1)\right\rceil$ qubits. As a natural suggestion we define
the states as binary notation of the number $k$. E.g. for every
$k$, the state $\left\vert B\right\rangle _{k}$ will have excited
those qubits, which stand on positions, on which in the binary
notation of the number $k$ is a $1$. On all other positions the
qubits will be in the ground state. The state $\left\vert
B\right\rangle _{k}$ will have the form
\begin{equation}
\left\vert B\right\rangle _{k}=\left\vert 0\right\rangle
^{\otimes(N-\left\lceil \log(N+1)\right\rceil)}\left\vert
b\right\rangle _{k},
\end{equation}
where $\left\vert b\right\rangle _{k}$ is a state of $\left\lceil
\log(N+1)\right\rceil$ qubits. After the whole procedure, we can
simply discard most of the qubits and keep only a logarithmic
number of them, still keeping the whole information.

Now the main task is to perform the transformation efficiently,
e.g. with at most polynomial number of elementary gates. This
seems not to be a crucial problem, as we will work strictly in the
computational basis, i.e. perform only transformations from one
basis state to other basis state. Similarly to the previous
transformation, we will perform it consecutively from the first to
the last qubit. First of all, let us remark that for $k<3$ the
transformation is trivial and no action is needed. The first
non-trivial number is $k=3$
where we need to transform $\left\vert 0\right\rangle ^{\otimes(N-3)}%
\left\vert 100\right\rangle \longrightarrow\left\vert
0\right\rangle ^{\otimes(N-3)}\left\vert 011\right\rangle .$ This
can be done easily by performing two C-NOT gates with the third
qubit as control and the first and second qubit as targets. After
that, we can perform a Toffoli gate with the first and second
qubits as controls and third qubit as target. Obviously, these
gates will act nontrivially only on the desired state, as all
other states $\left\vert C\right\rangle _{k}$ with $k\neq3$ have
$\left\vert 0\right\rangle $ on the third position. All states
with $k\neq3$ do not have $\left\vert 1\right\rangle $ both on
first and second position.

For $k>3$ we will perform similar operations. For every $k$ we
will perform C-NOT gates with the $k$th qubit as control and those
qubits as targets, which represent the number $k$ in binary
notation. At the end we will perform a single Toffoli gate with
all these (target) qubits as control, all other qubits on
positions smaller than $k$ as reversed controls (initiating the
operation if in the state $\left\vert 0\right\rangle $) and the
$k$th qubit as target. If we perform these operations subsequently
from smaller to bigger $k$ (from $3$ to $N$), they will always act
nontrivially only on the relevant state $\left\vert C\right\rangle
_{k}$.

For every $k$, we will need to perform at most $\log(k)$ C-NOT
gates and one Toffoli gate with $\log(k)$ controls. Such a Toffoli
gate can always be performed with quadratic number of C-NOT gates
\cite{Kniznica} with respect to the number of control qubits. So,
for every $k$, we need roughly $\log^{2}(k)$ C-NOT gates. Thus for
the whole transformation we will need no
more gates that in the order of%
\[
\sum_{k=3}^{N}\log^{2}(k)<\sum_{k=3}^{N}\log^{2}(N)<N\log^{2}(N)
\]
C-NOT gates.

\begin{figure}
\includegraphics[width=\hsize]{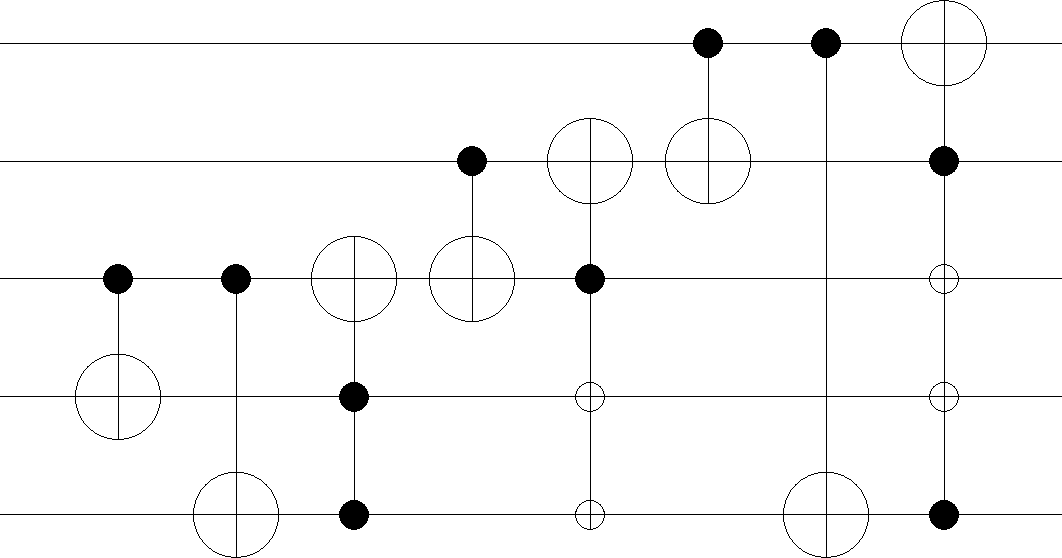}
\caption{ \label{figure2} Sequence of gates for the final step
operation in the five-qubit example.}
\end{figure}

\section{Noise analysis}
To show the capabilities of compressed state to resist to specific
noise, we have performed analysis on a specific noise model.
Within this model, every qubit is unitarily rotated by a specific
angle $\phi$ around a defined axis on the Bloch sphere. Such noise
can be imagined to be active, e.g. a magnetic field causing
precession of the stored (or sent) qubits. In the same way a
passive "noise" can be imagined, causing rotation or misalignment
of the reference frames.

We consider two scenarios. In first scenario, all $N$ qubits are
stored without compression and noise is acting an all the qubits.
In second scenario we first compress the $N$ qubits and store only
the non/trivial part of the state. The noise is acting now only on
the stored qubits. At the end we add qubits in the state
$\left\vert 0\right\rangle$ and decompress the state.

The decompression procedure is fully defined only for $N=2^{k}-1$
for every $k>0$. In other cases, the Hilbert space of the
compressed system has dimensions not used for storing information,
the unperturbed compressed state has zero amplitudes within this
dimensions. However, the noise can rotate the compressed state
such that also these dimension are used and in such a case one
would have to define the decompressing operation further to cover
the whole Hilbert space of compressed state.

\subsection{Global state fidelity}

\begin{figure}
\includegraphics[width=\hsize]{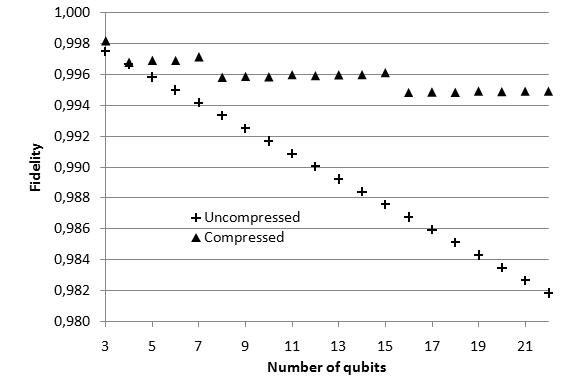}
\caption{ \label{figure3} Fidelity of global state after the
action of noise with and without compression and decompression.
The fidelity of the compressed state is clearly higher than the
fidelity of the uncompressed one.}
\end{figure}

Fidelity of the global state (\ref{stavPSI}) between the original,
unperturbed state with the state after action of noise an all
qubits is compared to the fidelity of state after compression,
action of noise and decompression. We average over all possible
input states of qubits and over all axis of rotation of the noise.
The results for $\phi=0.1 rad$ and different number of qubits are
shown in the Figure (\ref{figure3}). In this case the dimensions
of the Hilbert space of compressed state not used for storing
information will never contribute the the fidelity and therefore
we do not have to further define the decompression operation.

On the figure a clear structure is seen for the compressed state
with maximums of fidelities for specific number of qubits
(3,7,15). These are numbers for which the whole Hilbert space of
compressed state is used to store information. By increasing the
number of qubits, a sudden drop of fidelity appears due to
increase of the number of qubits of the compressed state, which
are subject to the action of noise. In any situation the fidelity
of the state after compression-decompression procedure is higher
than in the naive scenario of storing all qubits.

\subsection{Single qubit fidelity}

Here the fidelity of the single qubit state is examined under the
scenarios described above (with and without compression). In this
case the unused dimensions in the Hilbert space of compressed
state play may contribute to the result, therefore we examined a
specific case of $N=7$, where this is not the case. The symmetry
of the operation as well as of the errors guarantee the symmetry
of the resulting state. In general the state after decompression
will be entangled, resulting in mixed one qubit states, but still
all of them identical.

\begin{figure}
\includegraphics[width=\hsize]{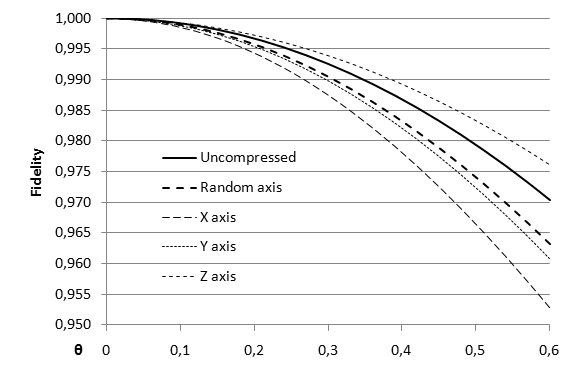}
\caption{ \label{figure4} Fidelity of one qubit state after the
action of noise with and without compression and decompression.
For compression scenario, results for noise acting around $x$, $y$
and $z$ axes, as well as for noise averaged through all axes are
shown. }
\end{figure}

The results o the calculations are shown in the Figure
\ref{figure4}) for different values of $\phi$. Results are
averaged through all input states. For uncompressed state the
resulting fidelity is not dependent on the axis of rotation of the
error. However, this is not the case for the compressed state,
therefor results for three specific axes of rotation, as well as
the result after averaging over all possible axes is shown.

We can conclude that in general the modelled type of noise is more
harmful to stored qubits. However, as only a small amount of
qubits is stored in the compressing scenario in comparison the
naive scenario, one can expect the ability to guarantee smaller
average errors. Even with the same error rate, we can obtain
better fidelity in the compressing scenario. If we have a
prediction about one more-stable axis, we can choose this to be
the $z$ axis of the compression-decompression operation (defining
the computational base and C-NOT operation). For errors causing
rotation around this axis the compressed state is more stable then
the uncompressed one.

\section{Conclusions} In this paper we have suggested a
\em{quantum} \em{compression} scheme for transformation of an
$N$-fold product state of a single qubit state into a state, which
is non-trivial only on $\left\lceil \log(N+1)\right\rceil$ qubits.
The same procedure also describes the inverse operation
(decompression). Both of these are effective in a sense that only
$O(N^{2})$ C-NOT gates are needed to perform the operations.

Possible use of the scheme is a quantum memory. Having more copies
of a single-qubit state, it might be very reasonable to compress
them into a state of only a few qubits, which will be more easily
protected against decoherence. If the copies are needed again, we
perform the decompression transformation.

In fact, if the stored state is exposed to errors causing a
rotation of the basis, a loss of fidelity is observed. If we
compare the scenario of storing uncompressed qubit states with the
scenario of storing the compressed state, the error (expressed in
the loss of fidelity) is significantly smaller in the latter case
\cite{Errors}. This is true even for big errors, where standard
error-correcting procedures fail.

The scenario of storing quantum information is imaginable e.g. in
a case when a single-qubit state is a result of a stage of quantum
computation and is needed as an input for a following stage of the
computation. If some stages of the computation can not be
performed immediately after each other (they may use the same
``hardware" which needs to be adjusted etc.), the $N$-fold
symmetric state of a single-qubit state (obtained after $N$ runs
of the computation) may be compressed and stored effectively, e.g.
with exponentially smaller memory demands and lower error rate, in
the meantime.

Another possible application is the sending of information about a
direction using quantum states. In cases when two communicating
parties share a reference frame, states resulting from the
suggested compression are very effective in communicating the
direction. If the sender has an option to send at most $n$ qubits,
he prepares a $2^{(n-1)}$-fold symmetric state of a single-qubit
state pointing in the desired direction. After compression, the
resulting compressed state will span the Hilbert space of exactly
$n$ qubits and can be sent to the receiver. He will now decompress
it back into $2^{(n-1)}$-fold symmetric state of a single-qubit
state and perform standard state tomography.

To compare the power of the suggested compression scheme with
known procedures, fidelities of sending of a direction via a
quantum channel using a small number of qubits are presented in
the Table \ref{Tabulka}. For big number of qubits, the fidelity of
our procedure grows as $F=1-\frac{1}{2^{n}+2}$, which is
exponentially faster than $F\sim1-\frac{\xi}{n^{2}}$ for the
scheme presented in \cite{Bagan} or for the case of sending simple
copies of the qubit state, where $F=$ $1-\frac{1}{n+2}$
\cite{Fidelity}. Thus by utilizing a shared reference frame
between communicating parties and paying the cost of it we can
gain an exponential decrease of fidelity loss.

\begin{table}[h!]
\begin{center}
\begin{tabular}
[c]{|l|l|l|l|l|l|l|}%
\hline
n & 1 & 2 & 3 & 4 & 5 & 6\\
\hline
 $\left\vert \psi\right\rangle ^{\otimes n}$ & $0.666$ &
$0.750$ & $0.800$ &
$0.833$ & $0.855$ & $0.875$\\
\hline
EB & $0.666$ & $0.789$ & $0.845$ & $0.911$ & $0.931$ & $0.943$\\
\hline
PB & $0.666$ & $0.800$ & $0.889$ & $0.941$ & $0.970$ & $0.992$\\
\hline
\end{tabular}
\end{center}
\caption{The comparison of fidelities of transfer of a direction
using quantum states in cases of a na\"\i ve scenario -- transfer
of multiple copies of a single qubit, the Bagan et. al. scheme [3]
(EB) and our compression scheme (PB).} \label{Tabulka}
\end{table}


\textbf{Acknowledgments.} This work  was supported by the Slovak
Research and Development agency project APVV-0673-07. We thank
Michal Sedl\'ak for helpful discussions  and Marcela Hrd\'a for
numerical analysis. MP would like to thank Action Austria-Slovakia
for support.



\begin{thebibliography}{00}

\bibitem[1]{Gisin}N. Gisin and S. Popescu, PRL 83, 432 (1999)

\bibitem[2]{Massar}S. Massar, PRA 62, 040101(R), (2000)

\bibitem[3]{Bagan}E. Bagan, M. Baig, A. Brey, R. Munoz-Tapia, R. Tarrach, PRL
85, 5230 (2000)

\bibitem[4]{Nielsen}M. A. Nielsen, I. L. Chuang, \emph{Quantum Computation and
Quantum Information}, Cambridge university press (2000)

\bibitem[5]{Schur}D. Bacon, I. L Chuang, A. W. Harrow, quant-ph/0601001
(2006) and references therein

\bibitem[6]{Kniznica}A. Barenco et.al., PRA 52, 3457 (1995)

\bibitem[7]{Kaye1} P. Kaye and M. Mosca, quant-ph/0101009 (2001)

\bibitem[8]{Kaye2} P. Kaye and M. Mosca, quant-ph/0407102 (2004)

\bibitem[9]{Cosin}V. Shende, S. Bullock and I. Markov, IEEE Transactions on
Computer-Aided Design 25 no. 6, 1000 (2006)

\bibitem[10]{Optimalizacia}V. Bergholm, J. Vartiainen, M. Mottonen and
M.Salomaa, Phys. Rev. A 71, 052330 (2005)

\bibitem[11]{Sedlak}M. Sedlak and M. Plesch, CEJP,
6(1) 2008, 128-134 (2008)

\bibitem[12]{Errors} M. Plesch and M. Hrda, in preparation

\bibitem[13]{Fidelity} S. Massar and S Popescu, PRL 77, 1259
(1999)

\end{thebibliography}
\end{document}